\documentclass[english,a4paper,prb,preprint,showpacs,amsmath,amssymb,a4paper]{revtex4}
\usepackage[T1]{fontenc}
\usepackage[latin1]{inputenc}
\usepackage{float}
\usepackage{amsmath}
\usepackage{graphicx}
\usepackage{amssymb}

\makeatletter

\providecommand{\tabularnewline}{\\}

\usepackage{float}



\usepackage{babel}
\makeatother
\begin{document}

\title{Theoretical Analysis of Resonant Inelastic X-Ray Scattering Spectra
in $\mathrm{LaMnO_{3}}$}

\author{Taeko Semba$^{1}$, Manabu Takahashi$^{2}$, and Junichi Igarashi$^{1}$}

\affiliation{$^{1}$Faculty of Science, Ibaraki University, Mito, Ibaraki 310-8512, Japan\\
 $^{2}$Faculty of Engineering, Gunma University, Kiryu, Gunma 376-8515, Japan}

\begin{abstract}
We analyze the resonant inelastic x-ray scattering (RIXS) spectra
at the $K$ edge of Mn in the antiferromagnetic insulating manganite
$\mathrm{LaMnO_{3}}$. We make use of the Keldysh-type Green-function
formalism, in which the RIXS intensity is described by a product of
an incident-photon-dependent factor and a density-density correlation
function in the $3d$ states. We calculate the former factor using
the $4p$ density of states given by an ab initio band structure calculation
and the latter using a multi-orbital tight-binding model. The ground
state of the model Hamiltonian is evaluated within the Hartree-Fock
approximation. Correlation effects are treated within the random phase
approximation (RPA). We obtain the RIXS intensity in a wide range
of energy-loss $2$-$15$ eV. The spectral shape is strongly modified
by the RPA correlation, showing good agreement with the experiments.
The incident-photon-energy dependence also agrees well with the experiments.
The present mechanism that the RIXS spectra arise from band-to-band
transitions to screen the core-hole potential is quite different from
the orbiton picture previously proposed, enabling a comprehensive
understanding of the RIXS spectra. 
\end{abstract}

\pacs{78.70.En 75.47.Lx 71.28.+d 78.20.Bh}

\maketitle

\section{Introduction\label{sec:Introduction}}

Recently, the perovskite-type manganites showing the colossal magnetoresistance
(CMR) effect have attracted much attention because the CMR effect
may be the key for high-capacity magnetic storage and spintoronics
as the next generation electronic devices. It is widely recognized
that the spin, charge and orbital degrees of freedom of the $3d$
electrons play important roles in the CMR effect in the manganites.
In the undoped material $\mathrm{LaMnO_{3}}$, the crystal structure\cite{Elemans1971JSSC3,Ritter1997PRB56}
belongs to the $Pbnm$ space group below 780K, where each $\mathrm{MnO_{6}}$
octahedron is noticeably rotated, tilted, and distorted due to the
cooperative Jahn-Teller distortion (JTD).\cite{Kanamori1960} The
orbital degeneracy in the $e_{g}$ states is lifted by the JTD or
by the orbital exchange interaction similar to the superexchage for
spins,\cite{Kugel1972} forming an orbital-ordered state, in addition
to an A-type antiferromagnetic long-range order.\cite{Wollan1955PR100,Matsumoto1970JPSJ29.606}
With hole doping, a variety of spin, charge, and orbital ordered phases
appear. It is widely accepted that the CMR effect is a consequence
of the destruction of the orbital ordered state in the presence of
a magnetic field, which may break the subtle balance among the multiple
degrees of freedom.

Resonant inelastic x-ray scattering (RIXS) technique using the x-ray
tuned at the transition-metal $K$ edge has been recognized as a powerful
tool to investigate the charge and orbital degrees of freedom in transition-metal
compounds. The RIXS process is described as a second-order optical
process, in which a $1s$-core electron is excited to an empty $4p$
state by absorbing an incident photon, then charge excitations are
created in the $3d$ states to screen the core-hole potential, and
finally the photoexcited $4p$ electron recombines with the $1s$-core
hole by emitting a photon. In the final state, the charge excitations
are left behind in the $3d$ states. In contrast to the optical conductivity
measurement, it allows us to investigate the momentum dependence of
the excitations because the corresponding x-rays have wavelengths
of the same order of lattice spacing. The clear momentum dependence
have been observed in cuprates.\cite{Hasan2000,Kim2002,Suga2005,Lu2006,Ellis2008PRB77}
Note that electron energy loss spectroscopy can also detect the momentum
dependence, but it cruelly suffers from multiple scattering effects.

Several RIXS experiments have been carried out on $\mathrm{LaMnO_{3}}$
and the hole-doped compounds. \cite{Inami2003PRB67,Ishii2004PRB70,Grenier2005PRL94}
For the undoped case, the RIXS spectra show three noticeable features
at the energy loss $2.5$, $8$, and $11$ eV, when the incident photon
energy is tuned near the $\mathrm{Mn}$ $K$ edge.\cite{Inami2003PRB67}
These features are also observed in the hole-doped manganites. \cite{Ishii2004PRB70,Grenier2005PRL94}
Based on the theoretical analysis exploiting the Liouville operator
method,\cite{Kondo2001} Inami et al argued that the $2.5$-eV peak
arises from an orbital excitation across the Mott gap, which is driven
by the offdiagonal part of the Coulomb interaction between a photoexcited
$4p$ electron and an $e_{g}$ electron.\cite{Inami2003PRB67} This
assumption, however, seems unrealistic, because such a offdiagonal
Coulomb interaction is much smaller than the Coulomb interaction between
the $1s$ hole and $3d$ electrons. This mechanism could not explain
the origin of the $8$- and $11$-eV peaks either. The purpose of
this paper is to elucidate the origin of the observed features in
the RIXS spectra of $\mathrm{LaMnO_{3}}$ on the basis of the well
established assumption that the charge excitations are created in
the $3d$ states to screen the core-hole potential. To obtain the
comprehensive understanding of the spectra, we use a multi-orbital
tight-binding model involving all the $\mathrm{Mn}$ $3d$ orbitals
and $\mathrm{O}$ $2p$ orbitals, and take account of the crystal
distortion forming $Pbnm$ by varying the transfer energy between
the $3d$ and $2p$ orbitals.

We calculate the RIXS intensity with the use of the formula proposed
by Nomura and Igarashi (NI).\cite{Nomura2004,Nomura2005} The NI formula
of the RIXS spectra is a kind of extension of the resonant Raman theory
developed by Nozi\`{e}res and Abrahams\cite{Nozieres1974} on the
basis of the many-body formalism of Keldysh.\cite{Keldysh1965} The
formula has advantages that it can rather easily be applied to complicated
models including many orbitals and provides clear physical interpretations
of the RIXS spectra. This formula utilizes the Born approximation
to the core-hole potential, and divides the RIXS intensity into two
factors; one describes an incident-photon dependence and the other
is the density-density correlation function in the $3d$ states. Similar
formulas have been derived by using different methods.\cite{Abbamonte1999}
The NI formula has been successfully applied to the quasi-one-dimensional
cuprates $\mathrm{SrCuO_{3}}$\cite{Nomura2004}, $\mathrm{CuGeO_{3}}$,\cite{Suga2005}
two-dimensional cuprate $\mathrm{La_{2}CuO_{4}}$\cite{Nomura2005,Igarashi2006,Takahashi2008JPSJ77},
and the prototypical AFM insulator $\mathrm{NiO}$.\cite{Takahashi2007NiO}
In these studies, the calculation have been carried out at zero temperature;
the electronic structures in the AFM phase have been calculated within
the Hartree-Fock approximation (HFA). %using the $d$-$p$ model or the multiorbital tight-biding models 
It is known that the HFA works well for the description of electronic
structures in the AFM insulators. Two-particle correlations have been
taken into account within the random phase approximation (RPA). It
has been found that the RPA correction modifies strongly the spectral
shape as a function of energy loss, having led to a good agreement
with the experiments. With these successes, we may conclude that the
RIXS intensity arises from band-to-band transitions to screen the
core-hole potential in the intermediate state. Multiple-scattering
contributions due to the core-hole potential have been also investigated
in order to examine the validity of the Born approximation, because
the core-hole potential is not definitely weak.\cite{Igarashi2006}
Having evaluated the contributions by means of the time-representation
method by Nozi\`{e}res and De Dominicis, \cite{Nozieres1969_1097}
it was found that the contributions could be mainly absorbed into
the shift of the core-level energy with minor modifications of the
RIXS spectral shape.\cite{Igarashi2006} This result partly justifies
the use of the Born approximation.

In the present study of LaMnO$_{3}$, we treat the strong Coulomb
interaction between the $3d$ orbitals within the HFA; we obtain an
A-type AFM insulating solution with an energy gap $\sim1.0\,\mathrm{eV}$,
where the occupation on the $3z^{2}-r^{2}$-type orbital is larger
than that on the $x^{2}-y^{2}$-type orbital at each $\mathrm{Mn}$
site. This result corresponds well to the observed orbital ordered
state. Note that the band structure calculation with the local density
approximation (LDA) fails to reproduce the wide energy gap. We calculate
the density-density correlation function at zero temperature by using
the energy bands thus obtained and by treating the two-particle correlations
within the RPA. We calculate another factor, the incident-photon-dependent
factor, using the $4p$ density of states (DOS) obtained from the
ab-initio band structure calculation. Combining the two factors, we
finally obtain the RIXS spectra, which show good agreement with the
experiments. \cite{Inami2003PRB67,Ishii2004PRB70,Grenier2005PRL94}
We could assign the $2.5$-eV peak in the RIXS spectra as a function
of energy loss to the electron-hole excitation across the Mott gap
in local majority spin channel, and the $8$-($11$-)eV peak to the
transition from the occupied $e_{g}$ states strongly hybridized with
the $\mathrm{O}$ $2p$ states in the deep valence band to the unoccupied
$e_{g}$ states in the local majority (minority) spin channel. We
also make clear the origin of the incident-photon-dependence of the
spectra by examining the corresponding factor in our formula.

The present paper is organized as follows. In Sec. \ref{sec:multi-orbital-tight-binding},
we introduce the multiorbital tight-binding model. In Sec. \ref{sec:Electronic-Structure-within},
we discuss the electronic structure within the HFA in the AFM phase
of $\mathrm{LaMnO_{3}}$. In Sec. \ref{sec:Formula-for-RIXS}, we
briefly summarize the NI formula for the RIXS spectra. In Sec. \ref{sec:Calculated-Results},
we present the calculated RIXS spectra in comparison with the experiments.
The last section is devoted to the concluding remarks.

\section{Electronic Structure of $\mathrm{LaMnO_{3}}$}

\subsection{Multiorbital tight binding model \label{sec:multi-orbital-tight-binding}}

\begin{figure}[H]

\begin{centering}
\includegraphics[scale=0.8]{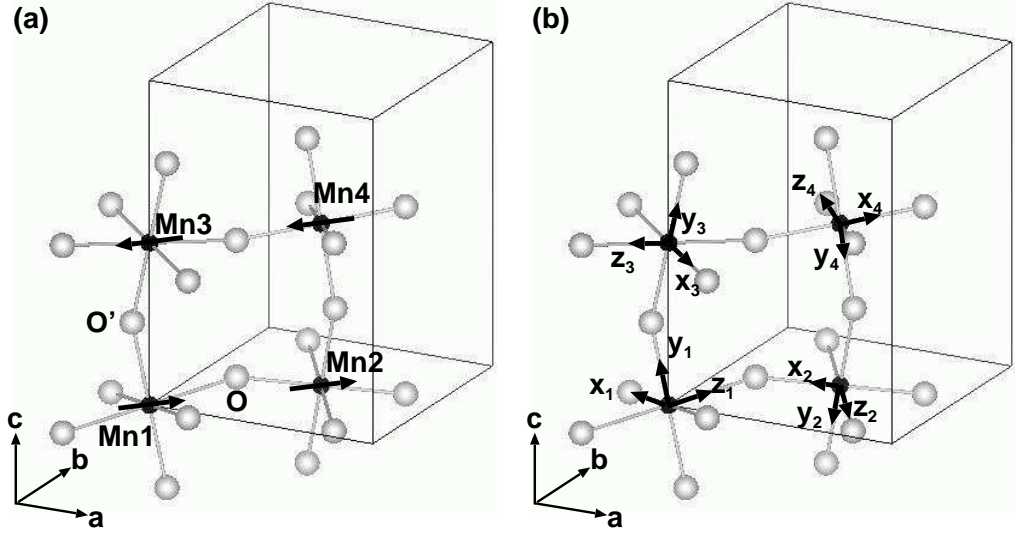} 
\par\end{centering}

\caption{(a) Sketch of the unit cell. $\mathrm{La}$ sites are omitted. The
experimentally observed crystal and magnetic structure is assumed.
The arrows on the $\mathrm{Mn}$ sites indicate the magnetic moment.
(b) Local coordinates at each $\mathrm{Mn}$ site. The $z_{\lambda}$
($\lambda=1,\cdots,4$) axis is taken to be parallel to the longest
$\mathrm{Mn}_{\lambda}$-$\mathrm{O}$ bond, while $x_{\lambda}$
and $y_{\lambda}$ axes are nearly parallel to the shortest $\mathrm{Mn_{\lambda}}$-$\mathrm{O}$
and the middle length $\mathrm{Mn_{\lambda}}$-$\mathrm{O}'$ bonds,
respectively.\label{fig:crys} }
\end{figure}

\begin{table}[H]

\caption{\label{table.1} Tight-binding parameters in units of eV. SK parameters
for the nearest $\mathrm{Mn}$-$\mathrm{O}'$ pair and for the nearest
$\mathrm{O}$-$\mathrm{O}'$ pair are given in the table, while the
other parameters are determined by adopting $\ell^{-\alpha}$ law
suggested by Harrison.\cite{Harrison_book1980} Slater Integral $F^{2}$
and $F^{4}$ are taken from ref. \onlinecite{Bocquet1992PRB46.3771}. }

\vskip10pt \begin{tabular}{cccccccc}
\hline 
$pd\sigma$ &
$-2.01$ &
&
$F^{0}$ &
$5.03$ &
&
$U$ &
$4.5$\tabularnewline
$pd\pi$ &
$0.93$ &
&
$F^{2}$ &
$9.74$ &
&
&
\tabularnewline
$pp\sigma$ &
$0.63$ &
&
$F^{4}$ &
$6.96$ &
&
&
\tabularnewline
$pp\pi$ &
$-0.18$ &
&
$E_{d}-E_{p}$ &
$-14.0$ &
&
$\Delta$ &
$4.0$\tabularnewline
\hline
\end{tabular}
\end{table}

We assume the crystal and magnetic structure of $\mathrm{LaMnO_{3}}$
as shown in Fig.~\ref{fig:crys}-a and also define the local coordinates
$x_{\lambda}$, $y_{\lambda}$, and $z_{\lambda}$ on the $\lambda$-th
$\mathrm{Mn}$ site in a unit cell as shown in fig. ~\ref{fig:crys}-b.
We introduce a tight-binding model involving all $\mathrm{Mn}$ $3d$
orbitals and $\mathrm{O}$ $2p$ orbitals. We exclude orbitals belonging
to the La atoms, since those orbitals play minor roles in the electronic
states near the insulating gap. Thus, the model Hamiltonian is expressed
as \begin{equation}
H=H_{0}+H_{\mathrm{I}},\label{eq:H}\end{equation}
 \begin{align}
H_{0} & =\sum_{im\sigma}E_{i}^{d}n_{im\sigma}^{d}+\sum_{jm\sigma}E_{j}^{p}n_{jm\sigma}^{p}\nonumber \\
 & +\sum_{\left(i,j\right)}\sum_{mm'\sigma}\left(t_{im,jm'}^{dp}d_{im\sigma}^{\dagger}p_{jm'\sigma}+\mbox{H.c.}\right)\nonumber \\
 & +\sum_{\left(j,j'\right)}\sum_{mm'\sigma}\left(t_{jm,j'm'}^{pp}p_{jm\sigma}^{\dagger}p_{j'm'\sigma}+\mbox{H.c.}\right),\label{eq:K}\\
H_{\mathrm{I}} & =\frac{1}{2}\sum_{i}\sum_{\nu_{1}\nu_{2}\nu_{3}\nu_{4}}g\left(\nu_{1},\nu_{2};\nu_{3},\nu_{4}\right)d_{\nu_{1}}^{\dagger}d_{\nu_{2}}^{\dagger}d_{\nu_{4}}d_{\nu_{3}}.\label{eq:G}\end{align}
 The part $H_{0}$ is the kinetic energy part, where $d_{im\sigma}$
and $p_{jm\sigma}$ denote the annihilation operators of an electron
with spin $\sigma$ in the $3d$ orbital $m$ at $\mathrm{Mn}$ site
$i$ and in the $2p$ orbital $m$ at $\mathrm{O}$ site $j$, respectively.
$n_{im\sigma}^{d}$ and $n_{jm\sigma}^{p}$ are the number operators
given by $d_{im\sigma}^{\dagger}d_{im\sigma}$ and $p_{jm\sigma}^{\dagger}p_{jm\sigma}$,
respectively. The transfer integrals, $t_{im,jm^{\prime}}^{dp}$ and
$t_{jm,j'm'}^{pp}$ are evaluated from the Slater-Koster (SK) two-center
integrals, $(pd\sigma)$, $(pd\pi)$, $(pp\sigma)$, $(pp\pi)$.\cite{Slater1954}
We neglect the hybridization between the $\mathrm{Mn}$ sites. In
order to take account of the JTD, we assume the $\ell^{-\alpha}$
law for the SK parameters suggested by Harrison,\cite{Harrison_book1980}
where $\ell$ represents the atomic distance between neighboring sites
and $\alpha=2$ for $pp\sigma$ and $pp\pi$, $\alpha=3.5$ for $pd\sigma$
and $pd\pi$. Thus, the JTD, rotation, and tilt of the $\mathrm{MnO_{6}}$
octahedron are incorporated into the model. The part $H_{\mathrm{I}}$
represents the intra-atomic Coulomb interaction on the $\mathrm{Mn}$
sites. The interaction matrix element $g\left(\nu_{1},\nu_{2};\nu_{3},\nu_{4}\right)$,
where $\nu$ stands for spin-orbit$\left(m\sigma\right)$, is written
in terms of the Slater integrals $F^{0}$, $F^{2}$, and $F^{4}$.
Among them, $F^{2}$ and $F^{4}$, which are known to be slightly
screened by solid-state effects, are taken from the cluster model
analysis of the x-ray photoemission spectroscopy.\cite{Bocquet1992PRB46.3771}
On the other hand, $F^{0}$ is known to be considerably screened,
so that we regard the value as an adjustable parameter. The Coulomb
interaction on $\mathrm{O}$ sites is absorbed into a renormalization
of the $\mathrm{O}$ $2p$ level parameters $E_{jm\sigma}^{p}$. The
$\mathrm{Mn}$ $d$-level position relative to the $\mathrm{O}$ $p$-levels
is given by the charge-transfer energy $\Delta$ defined as $\Delta=E_{d}-E_{p}+4U$
in the $d^{4}$ configuration, where $U$ is the multiplet-averaged
$d$-$d$ Coulomb interaction given by $U=F^{0}-\left(2/63\right)F^{2}-\left(2/63\right)F^{4}$.\cite{Bocquet1992PRB46.3771}
The charge transfer energy $\Delta$ is also treated as an adjustable
parameter in our calculation. The parameters used in the calculation
are listed in Table \ref{table.1}.

\subsection{Hartree-Fock approximation\label{sec:Electronic-Structure-within}}

\begin{figure}[H]

\begin{centering}
\includegraphics[clip,scale=0.55]{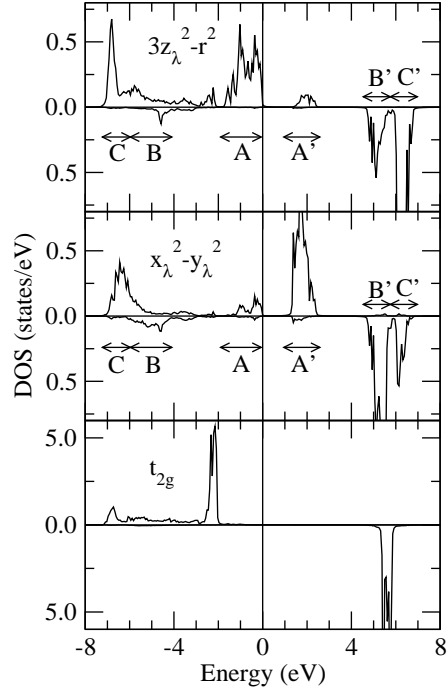} 
\par\end{centering}

\caption{\label{fig:ddos}DOS projected on the $\mathrm{Mn}$ $3z_{\lambda}^{2}-r^{2}$,
$x_{\lambda}^{2}-y_{\lambda}^{2}$, and $t_{2g}$ states. The origin
of energy is at the top of valence band. In each panel the upper half
represents the DOS for the local majority spin and the lower one for
the local minority spin.}
\end{figure}

\begin{figure}[H]

\begin{centering}
\includegraphics[clip,scale=0.55]{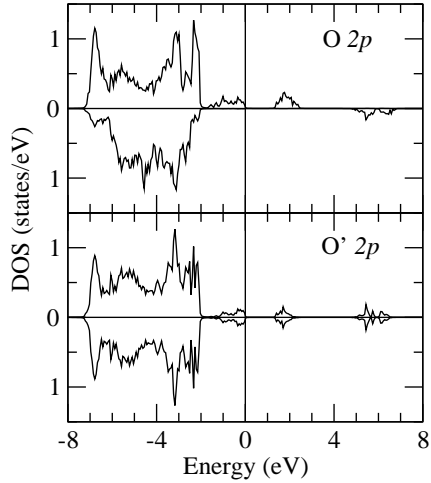} 
\par\end{centering}

\caption{\label{fig:pdos}DOS projected on the $\mathrm{O}$ $2p$ states
at the sites denoted by O and O' in fig. \ref{fig:crys}. The origin
of energy is at the top of valence band.}
\end{figure}

Assuming the A-type AFM order, we solve the tight-binding Schrodinger
equation within the HFA. We obtain a stable A-type AFM solution, which
has the energy gap $\sim1.0\,\mathrm{eV}$ and the spin moment at
$\mathrm{Mn}$ site $\sim2.0\,\hbar$. The densities of states (DOS)
projected onto the $3z_{\lambda}^{2}-r^{2}$, $x_{\lambda}^{2}-y_{\lambda}^{2}$,
and $t_{2g}$ states are shown in fig. \ref{fig:ddos}, and that projected
onto $2p$ states in fig. \ref{fig:pdos}. In the local coordinates,
they are independent of $\lambda$ ($1,\cdots,4$). The local majority
spin $3z_{\lambda}^{2}-r^{2}$ states are almost fully occupied and
mainly concentrate on the occupied energy ranges denoted by $\mathrm{A}$
and $\mathrm{C}$; they also concentrate on the unoccupied energy
range denoted by $\mathrm{A}'$. The local majority spin $x_{\lambda}^{2}-y_{\lambda}^{2}$
states are partially occupied on energy ranges denoted by $\mathrm{A}$
and $\mathrm{C}$; they also concentrate on the unoccupied energy
ranges denoted by $\mathrm{A}'$. On the other hand, the local minority
spin $e_{g}$ states highly concentrate on the unoccupied energy range
denoted by $\mathrm{B}'$ and $\mathrm{C}'$, although small amount
of them resides on the occupied energy range denoted by $\mathrm{B}$.
The difference in the occupation numbers between the $3z_{\lambda}^{2}-r^{2}$
and $x_{\lambda}^{2}-y_{\lambda}^{2}$ states implies that the $e_{g}$
states are orbitally ordered, corresponding to the experimentally
observed orbital order. In contrast to the $e_{g}$ states, the local
majority and minority spin $t_{2g}$ states are almost perfectly occupied
and unoccupied, respectively. The $e_{g}$ states around the energy
ranges denoted by $\mathrm{B}$ and $\mathrm{C}$ are highly hybridized
with $\mathrm{O}$ $2p$ states. In contrast, the occupied states
around the energy range denoted by $\mathrm{A}$ and unoccupied states
denoted by $\mathrm{A}'$ have very small weight of the $2p$ states.
This difference arises from the fact that the oxygen $2p$ states
are located in the relatively deep energy region as shown in fig.
\ref{fig:pdos}. LaMnO$_{3}$ is close not to the charge-transfer-type
insulator but to the the Mott-Hubbard-type insulator.

\begin{figure}[t]

\begin{centering}
\includegraphics{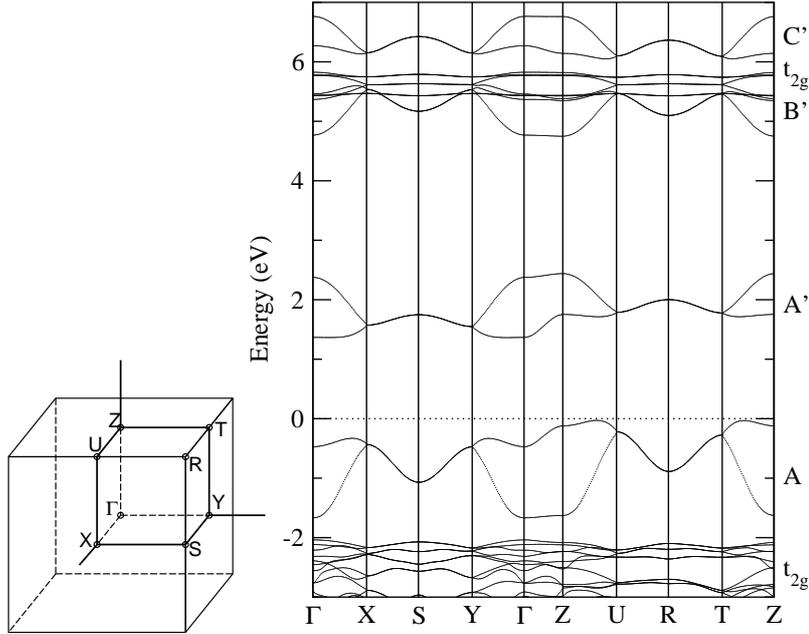} 
\par\end{centering}

\caption{\label{fig:disp}Dispersion curves along the several symmetric lines
vicinity of the gap. The origin of energy is at the top of valence
band.}
\end{figure}

The dispersion curves along some symmetric lines near the gap are
shown in fig. \ref{fig:disp}. Labels assigned to the curves correspond
to the states shown in the fig \ref{fig:pdos}. The curves corresponding
the states denoted by $\mathrm{A}$ and $\mathrm{A}'$ well reproduce
those calculated with the ab initio band structure calculation based
on the LDA$+U$ method\cite{Sawada1997} except for the magnitude
of the gap.

\section{RIXS process\label{sec:Formula-for-RIXS}}

We briefly summarize the NI formula for the RIXS, following ref. \onlinecite{Takahashi2007NiO}.
In the RIXS process, the incident photon is absorbed by exciting a
$\mathrm{Mn}$ $1s$ core electron to the unoccupied $\mathrm{Mn}$
$4p$ state, and a photon is emitted by recombining the $4p$ electron
and the core hole. This process may be described by \begin{equation}
H_{x}=w\sum_{{\bf q}\alpha}\frac{1}{\sqrt{2\omega_{{\bf q}}}}\sum_{i\eta\sigma}e_{\eta}^{(\alpha)}p_{i\eta\sigma}^{\prime\dagger}s_{i\sigma}c_{{\bf q}\alpha}{\rm e}^{i\mathbf{q}\cdot\mathbf{r}_{i}}+{\rm H.c.},\label{eq:Hx}\end{equation}
 where $w$ represents the dipole transition matrix element between
the $1s$ and the $4p$ states. We assume that $w$ is constant, since
it is expected to change little in the energy range of $20\,\mathrm{eV}$
above the absorption edge. It is known that the energy dependence
of $w$ gives minor effect on the $\mathrm{Cu}$ $K$-edge absorption
spectra.\cite{Igarashi2006} The $e_{\eta}$ represents the $\eta$-th
component ($\eta=x,y,z$) of the photon polarization vector. Annihilation
operators $p_{i\eta\sigma}^{\prime}$ and $s_{i\sigma}$ are for states
$4p_{\eta}$ and state $1s$ with spin $\sigma$ at $\mathrm{Mn}$
site $i$ , respectively. The annihilation operator $c_{{\bf q}\alpha}$
is for photon with momentum ${\bf q}$ and polarization $e_{\eta}^{(\alpha)}$.
In the intermediate state of the RIXS process, the core-hole potential
is acting on the $3d$ states, creating an electron-hole pair within
the Born approximation. The interaction is described as \begin{equation}
H_{1s-3d}=V\sum_{im\sigma\sigma'}d_{im\sigma}^{\dagger}d_{im\sigma}s_{i\sigma'}^{\dagger}s_{i\sigma'},\label{eq.h_1s3d}\end{equation}
 where $i$ runs over $\mathrm{Mn}$ sites. Note that, although the
value of the core-hole interaction $V$ is not known and may strongly
depend on the model, it is expected to be much larger than the $4p$-$3d$
offdiagonal Coulomb interaction. In the end of the process, an electron-hole
pair is left behind carrying momentum-energy $q\equiv\left(\mathbf{q},\omega\right)=\left(\mathbf{q}_{i}-\mathbf{q}_{f},\omega_{i}-\omega_{f}\right)$,
where $q_{i}=\left(\mathbf{q}_{i},\omega_{i}\right)$ and $q_{f}=\left(\mathbf{q}_{f},\omega_{f}\right)$
are momentum-energy of incident and scattered photons, respectively.

\begin{figure}[t]

\begin{centering}
\includegraphics[clip,scale=0.6]{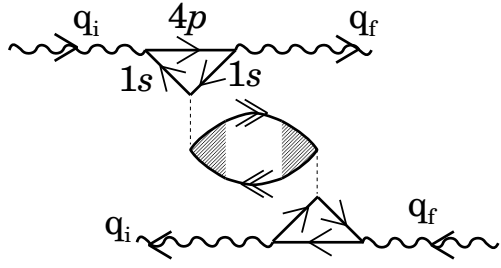} 
\par\end{centering}

\caption{\label{fig:diagram1}Diagram for the RIXS intensity within the Born
approximation for the $1s$ core-hole potential. The wavy and dotted
lines represent photon Green's functions and the core-hole interaction
$V$, respectively. The solid lines with the labels $4p$ and $1s$
represent the bare Green's functions for the $4p$ electron and the
$1s$ core electron, respectively. The elliptic part between the core-hole
interaction lines corresponds to the density-density correlation function
of the Keldysh type. The lines with double arrows are the Keldysh
type Green's functions. The shaded area represents the effective scattering
vertex renormalized by $3d$-$3d$ Coulomb interaction in the RPA. }
\end{figure}

The RIXS intensity is derived on the basis of the Keldysh-Green function
scheme. It is diagrammatically represented in Fig.\ref{fig:diagram1}.
Within the Born approximation for the core-hole potential, we obtain
\begin{equation}
W(q_{i},\mathbf{e}_{i};q_{f},\mathbf{e}_{f})=\frac{N\left|w\right|^{4}}{4\omega_{i}\omega_{f}}\sum_{\lambda m\sigma}\sum_{\lambda'm'\sigma'}e^{i\mathbf{q}\cdot(\mathbf{u}_{\lambda}-\mathbf{u}_{\lambda'})}Y_{\lambda m\sigma,\lambda'm'\sigma'}^{+-}(q)J_{\mathrm{B}\lambda\lambda'}\left(\omega_{i},\mathbf{e}_{i};\omega_{f},\mathbf{e}_{f}\right),\label{eq.general}\end{equation}
 where $\lambda$ indicates the $\lambda$-th $\mathrm{Mn}$ site
in a unit cell, $\mathbf{u}_{\lambda}$ represents the position vectors
of the $\lambda$-th $\mathrm{Mn}$ site in a unit cell. $N$ is the
number of unit cells. The factor $J_{\mathrm{B}\lambda\lambda'}\left(\omega_{i},\mathbf{e}_{i};\omega_{f},\mathbf{e}_{f}\right)$
describes the incident-photon dependence, which is given by \begin{equation}
J_{\mathrm{B}\lambda\lambda'}\left(\omega_{i},\mathbf{e}_{i};\omega_{f},\mathbf{e}_{f}\right)=\left(\sum_{\eta\eta'}e_{i\eta}L_{\mathrm{B}\lambda}^{\eta\eta'}\left(\omega_{i};\omega_{f}\right)e_{f\eta'}\right)\left(\sum_{\eta\eta'}e_{i\eta}L_{\mathrm{B}\lambda'}^{\eta\eta'}\left(\omega_{i};\omega_{f}\right)e_{f\eta'}\right)^{*},\label{eq:IPEF}\end{equation}
 where $e_{i\eta}$ ($e_{f\eta}$) is the $\eta$-th component of
the polarization vector $\mathbf{e}_{i}$ ($\mathbf{e}_{f}$) with
$\eta=x,y,z$, and $L_{\mathrm{B}\lambda}^{\eta\eta'}\left(\omega_{i};\omega\right)$
is given by

\begin{eqnarray}
L_{\mathrm{B}\lambda}^{\eta\eta'}\left(\omega_{i};\omega_{f}\right) & = & \frac{V}{N}\int_{\epsilon_{0}}^{\infty}\frac{\rho_{\lambda4p}^{\eta\eta'}\left(\epsilon\right){\rm d}\epsilon}{\left(\omega_{i}+\epsilon_{1s}+i\Gamma_{1s}-\epsilon\right)\left(\omega_{f}+\epsilon_{1s}+i\Gamma_{1s}-\epsilon\right)}.\label{eq.born}\end{eqnarray}
 The $\Gamma_{1s}$ represents the life-time broadening width of the
core-hole state, and the lower limit of the integral $\epsilon_{0}$
indicates the energy at the bottom of the $4p$ band. This expression
comes from the upper triangle in Fig.\ref{fig:diagram1}. The $\rho_{\lambda4p}^{\eta\eta'}$
is the DOS matrix in the $p$ symmetric states at the $\lambda$-th
$\mathrm{Mn}$ site, which may be given by \begin{equation}
\rho_{\lambda4p}^{\eta\eta'}\left(\epsilon\right)=\sum_{\sigma}\sum_{n\mathbf{k}}\phi_{\lambda\eta\sigma}^{*}\left(n,\mathbf{k}\right)\phi_{\lambda\eta'\sigma}\left(n,\mathbf{k}\right)\delta\left(\epsilon-\epsilon_{n}\left(\mathbf{k}\right)\right),\label{eq:4p_dos_mat}\end{equation}
 where $\phi_{\lambda\eta\sigma}\left(n,\mathbf{k}\right)$ is the
amplitude of $p_{\eta}$ component with spin $\sigma$ at the $\lambda$-th
$\mathrm{Mn}$ site in the band state specified by the band index
$n$ and crystal momentum $\mathbf{k}$ with energy $\epsilon_{n}\left(\mathbf{k}\right)$.
The factor $Y_{\lambda'm'\sigma',\lambda m\sigma}^{+-}\left(q\right)$
in Eq.~(\ref{eq.general}) is the density-density correlation function
of the Keldysh type, which is defined by %\begin{equation}
%Y_{d}^{+-}({\bf q},\omega)=\sum_{\lambda m\sigma}
%\sum_{\lambda^{\prime}m^{\prime}\sigma^{\prime}}
%Y_{\lambda m\sigma,\lambda^{\prime}m^{\prime}\sigma^{\prime}}^{+-}
%\left(q\right),\label{eq:Yd}
%\end{equation}
\begin{equation}
Y_{\lambda'm'\sigma',\lambda m\sigma}^{+-}({\bf q},\omega)=\int_{-\infty}^{\infty}\langle(\rho_{{\bf q}\lambda'm'\sigma'})^{\dagger}(\tau)\rho_{{\bf q}\lambda m\sigma}(0)\rangle{\rm e}^{i\omega\tau}{\rm d}\tau,\label{eq.y+-}\end{equation}
 where \begin{equation}
\rho_{\mathbf{q}\lambda m\sigma}=\sqrt{\frac{4}{N}}\sum_{{\bf k}}d_{\mathbf{k}+\mathbf{q}\lambda m\sigma}^{\dagger}d_{\mathbf{k}\lambda m\sigma},\label{eq:rho}\end{equation}
 with \begin{equation}
d_{{\bf k}\lambda m\sigma}=\sqrt{\frac{4}{N}}\sum_{n}d_{n\lambda m\sigma}{\rm e}^{i{\bf k\cdot r}_{n}}.\label{eq:fourier}\end{equation}
 The index $\lambda m\sigma$ specifies a tight-binding orbital at
site $\lambda$ with orbital $m$ and spin $\sigma$; $\lambda$ ($=1,2,3,4$)
is assigned to four $\mathrm{Mn}$ sites in a unit cell. Wavevector
$\mathbf{k}$ in eq.~(\ref{eq:rho}) runs over the first Brillouin
zone. Vector ${\bf r}_{n}$ in eq.~(\ref{eq:fourier}) represents
a position vector of the $n$-th unit cell. A single phase factor
${\bf k\cdot r}_{n}$ is assigned to all the $3d$ states in each
unit cell, and hence an extra factor ${\rm e}^{i{\bf q}\cdot({\bf u}_{\lambda}-{\bf u}_{\lambda'})}$
is required in eq.~(\ref{eq.general}).

We calculate the density-density correlation function (\ref{eq.y+-})
at zero temperature with taking account of the correlation effect
on the electron-hole pair by the RPA. Abbreviating the indices $\lambda m\sigma$
as $\xi$, it may be written as \begin{equation}
Y_{\xi'\xi}^{+-}\left(q\right)=\sum_{\xi'_{1}\xi'_{2}}\sum_{\xi_{1}\xi_{2}}\Lambda_{\xi'_{1}\xi'_{2},\xi'}^{*}\left(q\right)\Pi_{\xi'_{1}\xi'_{2},\xi_{1}\xi_{2}}^{+-\left(0\right)}\left(q\right)\Lambda_{\xi{}_{1}\xi{}_{2},\xi}\left(q\right),\label{eq:RPA}\end{equation}
 where\begin{eqnarray}
 &  & \Pi_{\xi_{1}\xi_{2},\xi'_{1}\xi'_{2}}^{+-(0)}\left(q\right)\nonumber \\
 & = & \frac{2\pi}{N}\sum_{{\bf k}}\sum_{j,j'}\delta\left(\omega-E_{j'}\left(\mathbf{k}+\mathbf{q}\right)+E_{j}\left(\mathbf{k}\right)\right)\left[1-n_{j'}\left(\mathbf{k}+\mathbf{q}\right)\right]n_{j}\left(\mathbf{k}\right)\nonumber \\
 & \times & \varphi_{\xi_{1},j'}\left(\mathbf{k}+\mathbf{q}\right)\varphi_{\xi'_{1},j'}^{*}\left(\mathbf{k}+\mathbf{q}\right)\varphi_{\xi'_{2},j}\left(\mathbf{k}\right)\varphi_{\xi_{2},j}^{*}\left(\mathbf{k}\right).\label{eq.keldysh1}\end{eqnarray}
 $E_{j}\left(\mathbf{k}\right)$ and $n_{j}\left(\mathbf{k}\right)$
are the eigen-energy and the occupation number of the eigenstate specified
by $j\mathbf{k}$, respectively. The $\varphi_{\lambda m\sigma,j}\left(\mathbf{k}\right)$
represents the amplitude of the $3d$ orbital and spin $m\sigma$
at the $\mathrm{Mn}$ site $\lambda$ in the energy eigenstate specified
by $j\mathbf{k}$ within the HFA. The RPA vertex $\Lambda_{\xi_{1}\xi_{2},\xi}(q)$
is given by\begin{equation}
\Lambda_{\xi_{1}\xi_{2},\xi}(q)=\left[\hat{I}-\hat{\Gamma}\hat{F}^{--}\left(q\right)\right]_{\xi_{1}\xi_{2},\xi\xi}^{-1},\label{eq.vertex}\end{equation}
 where $\hat{I}$ represents a unit matrix, and $\hat{\Gamma}$ is
the bare four-point antisymmetric vertex given by \begin{equation}
\left[\hat{\Gamma}\right]_{\xi_{1}\xi_{2},\xi_{3}\xi_{4}}=\left(g\left(\nu_{1}\nu_{2};\nu_{3}\nu_{4}\right)-g\left(\nu_{1}\nu_{2};\nu_{4}\nu_{3}\right)\right)\delta_{\lambda_{1}\lambda_{2}}\delta_{\lambda_{3}\lambda_{4}}\delta_{\lambda_{1}\lambda_{3}}.\label{eq:bare_four-point_vertex}\end{equation}
 The two-particle propagator $\hat{F}^{--}(q)$ is given by\begin{align}
\left[\hat{F}^{--}\left(q\right)\right]_{\xi_{1}\xi_{2},\xi_{3}\xi_{4}} & =\frac{1}{N}\sum_{jj'}\sum_{{\bf k}}\varphi_{\xi_{4},j}\left(\mathbf{k}\right)\varphi_{\xi_{2},j}^{*}\left(\mathbf{k}\right)\varphi_{\xi_{1},j'}\left(\mathbf{k}+\mathbf{q}\right)\varphi_{\xi_{3},j'}^{*}\left(\mathbf{k}+\mathbf{q}\right)\nonumber \\
 & \times\left[\frac{n_{j}\left(\mathbf{k}\right)\left[1-n_{j'}\left(\mathbf{k}+\mathbf{q}\right)\right]}{\omega-E_{j'}\left(\mathbf{k}+\mathbf{q}\right)+E_{j}\left(\mathbf{k}\right)+i\delta}-\frac{n_{j'}\left(\mathbf{k}+\mathbf{q}\right)\left[1-n_{j}\left(\mathbf{k}\right)\right]}{\omega-E_{j'}\left(\mathbf{k}+\mathbf{q}\right)+E_{j}\left(\mathbf{k}\right)-i\delta}\right].\label{eq:two_particle_propagator}\end{align}
 We evaluate the density-density correlation function (\ref{eq.y+-})
using eqs. (\ref{eq:RPA})--(\ref{eq:two_particle_propagator}). For
more details of the derivation, see refs.\onlinecite{Igarashi2006},
\onlinecite{Takahashi2007NiO}, and \onlinecite{Takahashi2008JPSJ77}.
Note that the terms involving $\Pi^{-+(0)}\left(q\right)$ are neglected;
they have no contribution for $\omega>0$ at zero temperature, because
$\Pi^{-+(0)}\left(q\right)\propto\sum_{\mathbf{k}}\left(1-n\left(\mathbf{k}\right)\right)n\left(\mathbf{k}+\mathbf{q}\right)\delta\left(\omega-E_{j'}\left(\mathbf{k}+\mathbf{q}\right)+E_{j}\left(\mathbf{k}\right)\right)$
with omitting unimportant factors.

\section{Calculated Results\label{sec:Calculated-Results}}

\begin{figure}[t]

\begin{centering}
\includegraphics[clip,scale=0.7]{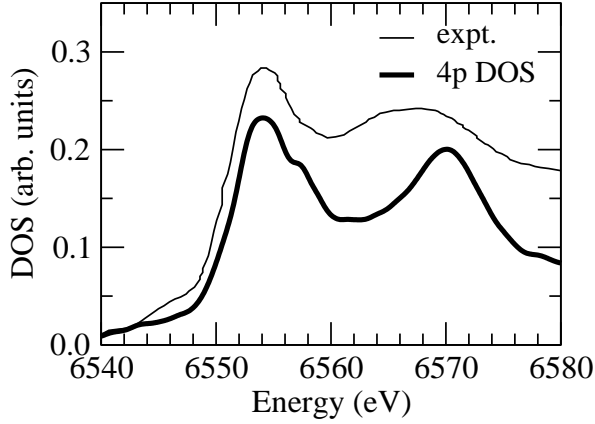} 
\par\end{centering}

\caption{\label{fig:dos-4p}The LDA $4p$ DOS convoluted by the Lorentzian
function with FWHM $2\Gamma_{1s}=2\,\mathrm{eV}$ (thick solid line
curve). In connection with the absorption coefficient, the origin
of energy is shifted so that the prominent peak locates at $6554\,\mathrm{eV}$.
Thin line curve is the absorption spectra reproduced from ref. \onlinecite{Subias_RPB56_8183}.}
\end{figure}

In order to calculate the incident-photon-dependent factor $J_{\mathrm{B}\lambda\lambda'}\left(\omega_{i},\mathbf{e}_{i};\omega_{f},\mathbf{e}_{f}\right)$,
we need the $4p$ DOS $\rho_{\lambda4p}^{\eta\eta'}(\epsilon)$ on
the $\lambda$-th $\mathrm{Mn}$ site. We evaluate the $4p$ DOS using
full-potential-linear-augmented-plane-wave band structure calculation
based on the LDA$+U$ method. It should be noted here that the $\mathrm{Mn}$
$4p$ DOS depends on $\eta$, $\eta'$ and on $\mathrm{Mn}$ site
$\lambda$ due to the strong JTD, giving rise to the $\mathrm{Mn}$
$K$-edge resonant elastic x-ray scattering intensity on forbidden
Bragg spots. \cite{Elfimov1999,Benfatto1999,Takahashi1999} Figure
\ref{fig:dos-4p} shows the $4p$ DOS averaged with $\eta$ ($\sum_{\eta}\rho_{\lambda4p}^{\eta\eta}(\epsilon)$)
and convoluted with a Lorentzian function of FWHM $2\Gamma_{1s}=2\,\mathrm{eV}$,
in comparison with the absorption experiment.\cite{Subias_RPB56_8183}
We set the energy difference between the $\mathrm{Mn}$ $1s$ level
and the prominent peak in the $4p$ DOS to be $6554\,\mathrm{eV}$.
Under the condition that the dipole matrix element is constant and
that the interaction is neglected between the core hole and the $4p$
electron, the $4p$ DOS becomes proportional to the $\mathrm{Mn}$
$K$-edge absorption spectra. The agreement with the experiment indicates
that the above condition is nearly satisfied.

\begin{figure}[t]

\begin{centering}
\includegraphics[clip,scale=0.7]{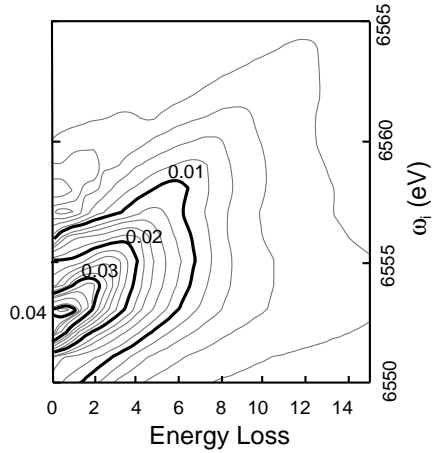} 
\par\end{centering}

\caption{\label{fig:LBC} Contour plot of $J_{\mathrm{A}}\left(\omega_{i};\omega_{f}\right)$
as a function of energy loss $\omega=\omega_{i}-\omega_{f}$ and incident
photon energy $\omega_{i}$. $2\Gamma_{1s}=2$ eV. Units of the intensity
are arbitrary.}
\end{figure}

Equations (\ref{eq.general})-(\ref{eq.born}) indicate that the photon
polarization dependence of the RIXS intensity correlates only with
the $\mathrm{Mn}$ $4p$ states. Therefore, the polarization dependence
would not have relevant information on the charge excitations in the
$3d$ states. Hence, we will not touch on the polarization dependence.
Assuming polarization-unresolved analysis, we neglect the offdiagonal
elements of the $4p$ DOS matrix and replace each diagonal element
with the averaged value for simplicity. Accordingly, the factor $J_{\mathrm{B}\lambda\lambda'}\left(\omega_{i},\mathbf{e}_{i};\omega_{f},\mathbf{e}_{f}\right)$
is replaced with the averaged site-independent factor $J_{\mathrm{A}}\left(\omega_{i};\omega_{f}\right)\sim\left|\sum_{\eta}L_{\mathrm{B}\lambda}^{\eta\eta}\left(\omega_{i};\omega_{f}\right)\right|^{2}$.
Figure \ref{fig:LBC} shows the contour plot of $J_{\mathrm{A}}\left(\omega_{i};\omega_{f}\right)$
as a function of incident photon energy $\omega_{i}$ and energy loss
$\omega=\omega_{i}-\omega_{f}$. The enhancement appears around the
energy loss of $2$ eV for the incident photon energy around $6554$
eV. As the incident photon energy increases, the enhancement peak
moves toward higher energy loss region with decreasing the intensity.
It is expected that the factor $J_{\mathrm{B}\lambda\lambda'}\left(\omega_{i},\mathbf{e}_{i};\omega_{f},\mathbf{e}_{f}\right)$
has dependence similar to the factor $J_{\mathrm{A}}\left(\omega_{i};\omega_{f}\right)$,
although the enhancement peak position and the intensity may somewhat
depend on the photon polarization.

Another factor $Y_{\lambda'm'\sigma,\lambda m\sigma}^{+-}(q)$ mainly
determines the structure of the RIXS spectra as a function of energy
loss. We calculate this factor from eqs. (\ref{eq.y+-})--(\ref{eq:two_particle_propagator})
using the tight-binding wavefunction within the HFA and by including
the RPA correction. The transition from the occupied $e_{g}$ states
in the energy ranges denoted by $\mathrm{A}$, $\mathrm{B}$ and $\mathrm{C}$
to the unoccupied $e_{g}$ states in those denoted by $\mathrm{A}'$,
$\mathrm{B}'$, and $\mathrm{C}'$ can be driven due to the $1s$
core-hole potential, because the occupied states and the unoccupied
states comprise the $e_{g}$ states with the same local symmetry.
Therefore, the $e_{g}$ states can contribute to the RIXS intensity.
In contrast to the $e_{g}$ states, the $t_{2g}$ states cannot contribute
to the RIXS intensity, since the local majority and minority spin
$t_{2g}$ states are almost perfectly occupied and unoccupied, respectively,
and the spin-flip transition is not allowed.

\begin{figure}[H]
 \includegraphics[scale=0.5]{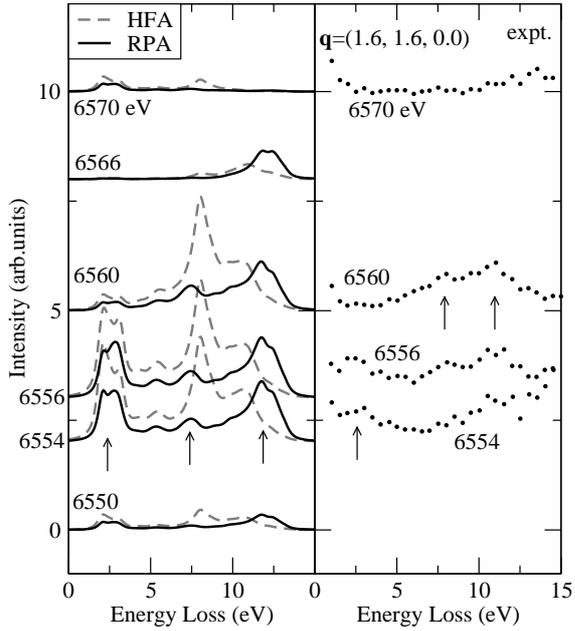}

\caption{\label{fig:RIXS161600} RIXS spectra at $\mathbf{q}=\left(1.6,1.6,0.0\right)$
as a function of energy loss for several incident photon energies.
Calculated RIXS spectra are shown in the left panel and the observed
spectra reproduced from ref. \onlinecite{Inami2003PRB67} in the
right panel. The solid and dashed curves are spectra calculated by
including the RPA correction and within the HFA, respectively. The
incident photon energies are shown besides each spectral curve in
units of eV. Arrows indicate characteristic features in the spectra.}
\end{figure}

\begin{figure}[H]
 \includegraphics[scale=0.5]{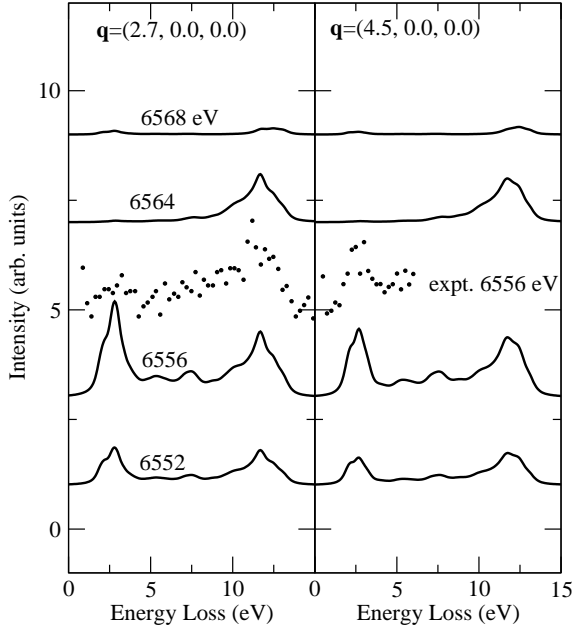}

\caption{\label{fig:RIXS27-47} RIXS spectra at $\mathbf{q}=\left(2.7,0.0,0.0\right)$
and $\left(4.5,0.0,0.0\right)$ as a function of energy loss for several
incident photon energies. %$Pbnm$ setting is assumed for the definition of ${\bf q}$.
The solid curves represent the spectra calculated by including the
RPA correction. Dotted ones are the observed spectra at $\hbar\omega_{i}=6556$
eV reproduced from ref. \onlinecite{Ishii2004PRB70}.}
\end{figure}

Combining two factors, we obtain the RIXS spectra. We convolute the
result with a Lorentzian function of $\mathrm{FWHM}=0.5\,\mathrm{eV}$
for taking account of the instrumental resolution. Figure \ref{fig:RIXS161600}
shows the spectra thus evaluated at the momentum transfer ${\bf q}=(1.6,1.6,0.0)$
as a function of energy loss for several incident photon energies
in comparison with the experiments. \cite{Inami2003PRB67,Ishii2004PRB70,Grenier2005PRL94}
We obtain continuous spectra ranging from $\omega=2$ eV to $15$
eV within the HFA; there are three features around $\omega=2.5$,
$7.5$, and $12\,\mathrm{eV}$, which arise from transitions of $\mathrm{A}\rightarrow\mathrm{A}'$,
$\mathrm{C}\rightarrow\mathrm{A}'$ in the local majority spin channel,
and $\mathrm{B}\rightarrow\mathrm{B}',\,\mathrm{C}'$ in the local
minority spin channel, respectively, within the $e_{g}$ states (see
Figs. \ref{fig:ddos} and \ref{fig:disp}). As already stated, the
$t_{2g}$ states could not contribute to the RIXS intensity. The spectral
shape within the HFA is drastically modified by the RPA correction;
the intensity around the energy loss $\omega\sim2.5$ and $8$ eV
is suppressed and that around $\omega\sim12$ eV is enhanced. The
spectral shape thus modified corresponds well to the experimentally
observed features at $\omega=2.5$, $8$, and $11$ eV. Another characteristic
is that the weight of the RIXS intensity moves toward higher energy
loss region as the incident photon energy increases. This change is
nicely reproduced by the calculation, mainly due to the effect of
the incident-photon-energy factor $J_{\mathrm{A}}$. Figure \ref{fig:RIXS27-47}
shows another comparison with the experiment at other momentum transfers
$\mathbf{q}=\left(2.7,0.0,0.0\right)$ and $\left(4.5,0.0,0.0\right)$.
\cite{Ishii2004PRB70} The calculated spectra agree with the experiment.

\begin{figure}[H]
 \includegraphics[scale=0.7]{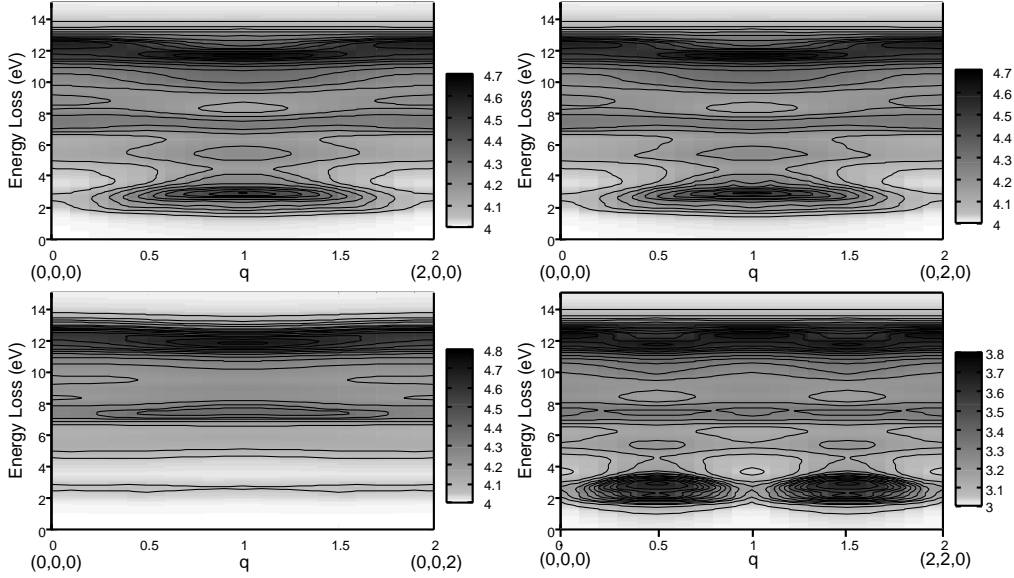}

\caption{RIXS spectra as a function of momentum transfer $\mathbf{q}$ along
$\left(0,0,0\right)$-$\left(2,0,0\right)$, -$\left(0,2,0\right)$,
-$\left(0,0,2\right)$, and -$\left(2,2,0\right)$ lines. $Pbnm$
setting is assumed for the definition of ${\bf q}$. The incident-photon
energy is fixed at $\omega_{i}=6556$ eV. The unit of the intensity
is arbitrary. \label{fig:RIXS-Q-dep}}
\end{figure}

Figure \ref{fig:RIXS-Q-dep} shows a contour plot of RIXS spectra
as a function of energy loss and momentum transfer ${\bf q}$ along
$\left(0,0,0\right)$-$\left(2,0,0\right)$, -$\left(0,2,0\right)$,
-$\left(0,0,2\right)$, and -$\left(2,2,0\right)$ lines. The incident
photon energy $\omega_{i}$ is fixed at $6556$ eV. Three ridges around
the energy loss $2.5$, $7.5$, and $12$ eV correspond to the three
peaks around the energy loss $2.5$, $7.5$, and $12$ eV, respectively,
in figs. \ref{fig:RIXS161600} and \ref{fig:RIXS27-47}. The peak
\textit{position} only slightly depends on the momentum transfer $\mathbf{q}$,
moving within $\sim1$ eV. This is consistent with the experiments
and may be due to the fact that the states relevant to the RIXS process
concentrate on the narrow energy ranges. Note that the peaks at $2.5$,
$7.5$, and $12$ eV correspond to transitions $\mathrm{A}\rightarrow\mathrm{A}'$,
$\mathrm{C}\rightarrow\mathrm{A}'$ in the local majority spin channel,
and $\mathrm{B}\rightarrow\mathrm{B}',\,\mathrm{C}'$ in the local
minority spin channel, respectively (see Figs. \ref{fig:ddos} and
\ref{fig:disp}). In contrast, the peak \textit{intensity} noticeably
depends on the momentum transfer showing the period of $2$ along
the $a$, $b$, and $c$ directions, although this fact has not been
emphasized in the experiments.\cite{Inami2003PRB67,Ishii2004PRB70,Grenier2005PRL94}
It does not seem easy, however, to explain the origin of this dependence,
since the intensities depend sensitively on the $3d$ weights in the
energy bands (see Eq. (\ref{eq.keldysh1})). Different from the present
case, the peak positions as a function of energy loss have been found
clearly moving with varying momentum transfer in $\mathrm{La_{2}CuO_{4}}$
in the high-resolution experiments. \cite{Hasan2000,Kim2002,Suga2005,Lu2006,Ellis2008PRB77}
It should be noted here that the peak shift with varying momentum
transfer should not be interpreted as a dispersion relation of a kind
of exciton but as a change of spectral weight in the continuum spectra.
The crystal momentum dependence of the weight of the states relevant
to the RIXS process and the energy dispersion of the single electron
states determine the momentum transfer dependence of the RIXS intensity.

\section{Concluding Remarks\label{sec:Concluding-Remarks}}

We have analyzed the incident-photon-energy and momentum dependence
on the RIXS spectra as a function of energy loss in $\mathrm{LaMnO_{3}}$.
We have utilized the formula developed by Nomura and Igarashi, which
expresses the RIXS spectra by a product of the photon dependent factor
and the density-density correlation function. The former factor, which
describes the dependence on the incident-photon energy and polarization,
has been calculated by using the $\mathrm{Mn}$ $4p$ DOS given by
the ab initio band structure calculation. The latter function describes
the charge excitations generated in the $3d$ states to screen the
core-hole potential in the intermediate state.

Having combined the above factors, we have calculated the RIXS spectra
as a function of energy loss. We have obtained a continuous spectrum
consisting of three features in good agreement with the experiments.
We have found that the RPA correlation modifies drastically the spectral
shape, which indicates the importance of electron correlations for
making quantitative analysis. We have demonstrated that the peak intensity
depends strongly on the momentum transfer $\mathbf{q}$ with the periodicity
$2$ along the $a$, $b$, and $c$ directions while the three peak
positions as a function of energy loss only slightly depend on $\mathbf{q}$.
Furthermore, we have obtained the spectral weight moving toward high
energy-loss region with increasing incident-photon energy, in agreement
with the experiments.

The present analysis naturally leads to a RIXS picture that the spectra
are brought about by band-to-band transitions (augmented by the RPA
correction) in order to screen the core-hole potential. This picture
is different from a previous picture that the the spectra are brought
about by $3d$ orbital excitations (orbitons) created with the use
of the offdiagonal part of the Coulomb interaction between the photo-excited
$4p$ electron and $3d$ electrons.\cite{Kondo2001,Inami2003PRB67}
This {}``orbiton\char`\"{} picture seems unreasonable, since the
$4p$-$3d$ offdiagonal Coulomb interaction, which causes orbitons,
is much smaller than the $1s$-$3d$ Coulomb interaction, which causes
band-to-band transitions. Actually the {}``orbiton\char`\"{} picture
was not successful in providing the spectra comparable to the experiment.\cite{Inami2003PRB67}
Note that the two pictures give different selection rules for the
creation of excitation; the present picture forbids the transition
from $3z^{2}-r^{2}$-type orbital to the $x^{2}-y^{2}$-type orbital,
while the latter allows it by changing $4p$ states.

For doped cuprates and manganites, experimental data have been accumulated.\cite{Ishii2004PRB70,Grenier2005PRL94}
In doped cuprates the RIXS spectra have been analyzed by the same
formula as the present one within the HFA.\cite{Markiewicz2006} As
already demonstrated in Refs. \onlinecite{Nomura2004,Nomura2005,Igarashi2006,
Takahashi2007NiO,Takahashi2008JPSJ77}, the HFA alone, without taking
account of the RPA correlation, is not sufficient for quantitative
understanding of the spectra even in the undoped materials. In doped
materials, electron correlations may become more important, and the
HFA-RPA scheme would not work well. Also, the Born approximation may
be insufficient for treating the core-hole potential, since many electron-hole
pairs could be created in the absence of the energy gap. It seems
hard to answer these questions by analyses with a detailed model like
the present paper, and such studies are left in future.

This work was partially supported by a Grant-in-Aid for Scientific
Research from the Ministry of Education, Culture, Sports, Science,
and Technology, Japan.

\bibliographystyle{apsrev} \bibliographystyle{apsrev}
\bibliography{Bibfile_new}

\end{document}